\pdfoutput=1
\documentclass[english]{article}
\usepackage[T1]{fontenc}
\usepackage[latin9]{inputenc}
\usepackage{amssymb}
\usepackage{graphicx}

\makeatletter

\providecommand{\tabularnewline}{\\}

\usepackage{lineno}

\makeatother

\usepackage{babel}
\begin{document}

\title{Development of an electrical model of a resistive micromegas}

\author{Jérôme SAMARATI}

\maketitle

\section{Model description}

We have developped a model to simulate the behavior of micromegas
(MICROMEsh GAseous Structure) \cite{ref:Gio}  geometry to a discharge using an electronic
software (Virtuoso \cite{ref:Virt}).

The principle of operation of a micromegas detector is presented in
figure \ref{fig1:Schema_Principe_Micromegas_with_Strips}.

\begin{center}
\begin{figure}[h]
\begin{centering}
\includegraphics[scale=0.3]{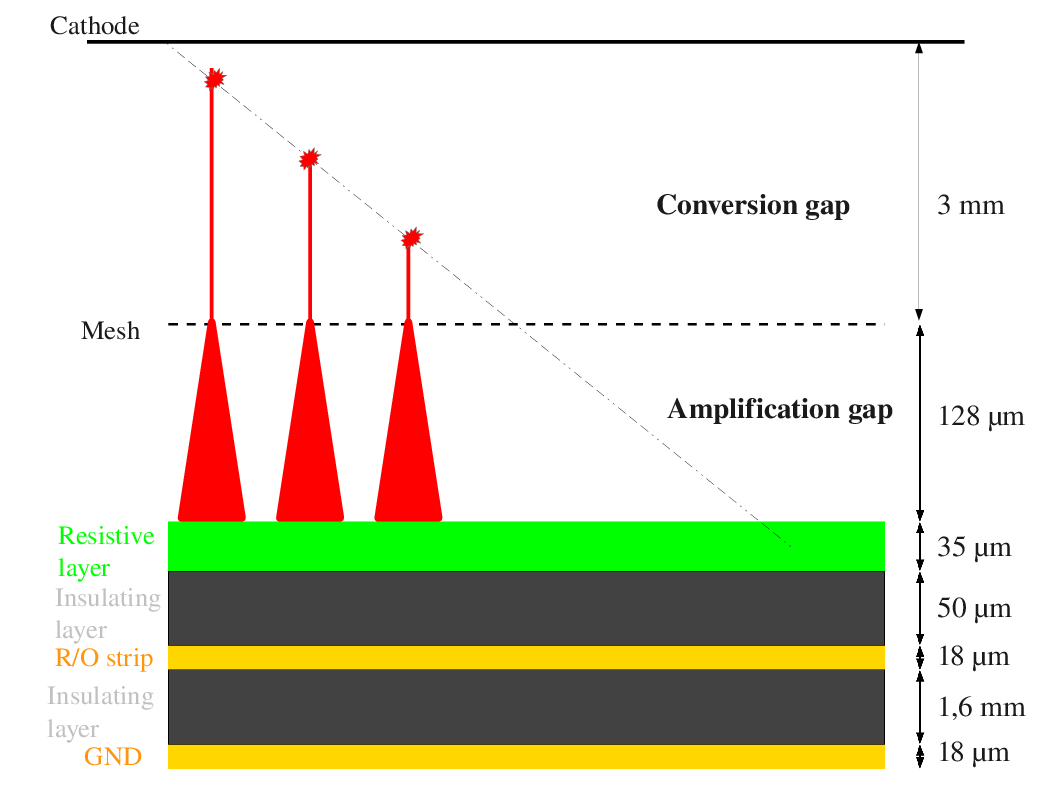}\caption{\label{fig1:Schema_Principe_Micromegas_with_Strips}Principle of operation
of a micromegas detector}

\par\end{centering}

\end{figure}

\par\end{center}

When a charged particle crosses the sensitive volume, it creates primary
electrons in the conversion gap whose drift toward the mesh under
the influence of a low electric field (few kV/cm). Due to high field
ratio between the conversion gap and the amplification gap, most of
them enter the amplification gap and are multiplicated under a high
electric field, $E\sim30$ kV/cm, due to the avalanche mechanism.
The motion of the charges in the amplification gap induces a signal
on the anode plane made of strips of 120 mm length. 

In our simulations, we have described the detector by its electrical
parameters as presented in figure \ref{fig2:Elementary_Cell_100um}.
The simulation has been done for different values of resistivity ($10k\Omega/\square$,
$100k\Omega/\square$,$1M\Omega/\square$ and$10M\Omega/\square$)
and for 3 different widths of strips ($100$, $200$ and $300\mu m$).
For practical reasons, we have segmented the strips according to its
width in order to keep a square elementary cell ($100\times100\mu m^{2}$,
$200\times200\mu m^{2}$ and $300\times300\mu m^{2}$) to match the
unit of resistivity.

\begin{center}
\begin{figure}[h]
\begin{centering}
\includegraphics[scale=0.3]{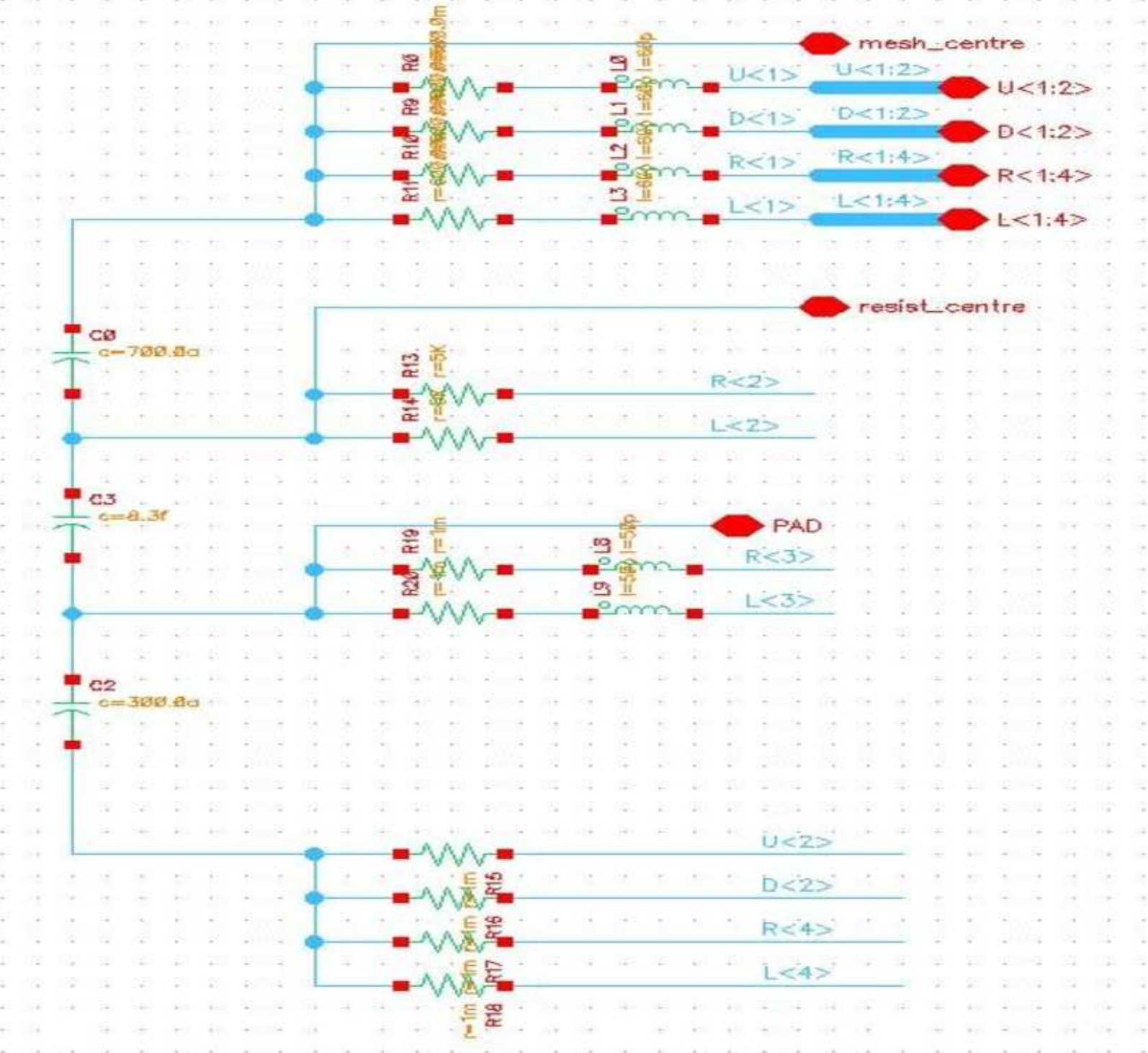}
\par\end{centering}

\caption{\label{fig2:Elementary_Cell_100um}Elementary cell for a $100\mu m$
width strip}

\end{figure}

\par\end{center}

The mesh and the strip are described by a pattern of inductances ($L_{mesh}$
and $L_{strip}$) and resistors, and the resistive layer and the ground
plane are described by a serie of resistance. We have introduced three
capacitances to modelise the capacitance of the gas gap ($C_{gas}$),
the capacitance of the insulating layer ($C_{dielectric}$) between
the resistive strip and the read out one and the capacitance of the
insulating layer between the read out strip and the ground plane ($C_{gnd}$).
The values of those parameters are presented in table \ref{tab5: Values of simulation parameters }.

\begin{center}
\begin{table}[h]
\begin{centering}
\begin{tabular}{|c|c|c|c|}
\hline 
 & \multicolumn{3}{c|}{Strip width}\tabularnewline
\hline 
\hline 
 & $100\mu m$ & $200\mu m$ & $300\mu m$\tabularnewline
\hline 
$C_{gas}$ (fF) & 0.7 & 2.8 & 6.3\tabularnewline
\hline 
$C_{dielectric}$ (fF) & 8.32 & 33 & 75\tabularnewline
\hline 
$C_{gnd}$ (fF) & 0.26 & 1 & 2.3\tabularnewline
\hline 
$L_{mesh}$ (pH) & 60 & 120 & 180\tabularnewline
\hline 
$L_{strip}$ (pH) & 48.5 & 83 & 112.5\tabularnewline
\hline 
\end{tabular}\caption{\label{tab5: Values of simulation parameters }Computed parameters
used in the model for the three widths of strip}

\par\end{centering}

\end{table}

\par\end{center}

In case of the mesh, the resisitive layer and the strip, we add an
ouput point in order to extract the useful parameters (voltage and
current).

The elementary cells are connected to their neighbors through two
directions (right and left) to modelise a complete strip of 120 mmm
length. It is important to notice that in our model, we have segmented
the strip with a pitch of $900\mu m$ for a strip of $100\mu m$ and
$300\mu m$ (121 elements which represents a total length of 108.9
mm) and with a pitch of 1 mm with a strip width of $200\mu m$ (121
elements which represents a total length of 121 mm). We have decided
to have an odd number of elementary cells in our pitch (9 for $100\mu m$,
5 for $200\mu m$ and 3 for $300\mu m$) and also in our total number
of pitch (121) in order to have a discharge located in the center
of the strip, to avoid any asymetry that could play a role in the
results. An example for a cell of $100\mu m$ is shown in figure \ref{fig3:1mm_Pitch_100um}.

\begin{center}
\begin{figure}[h]
\begin{centering}
\includegraphics[scale=0.3]{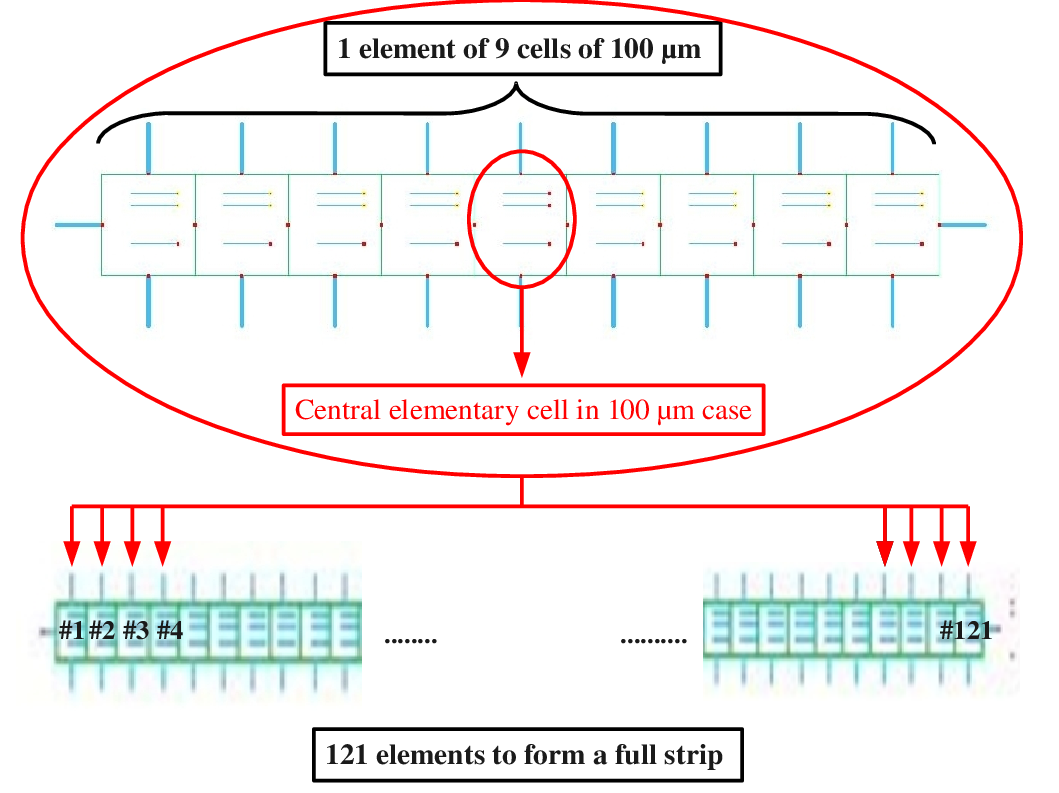}\caption{\label{fig3:1mm_Pitch_100um}Full strip made of 121 elements of 9
elementary cells of $100\mu m$}

\par\end{centering}

\end{figure}

\par\end{center}

We have done the simulation for a single strip geometry as it shown
on figure \ref{fig4:Simulation_1_Strip}.

\begin{center}
\begin{figure}[h]
\begin{centering}
\includegraphics[scale=0.3]{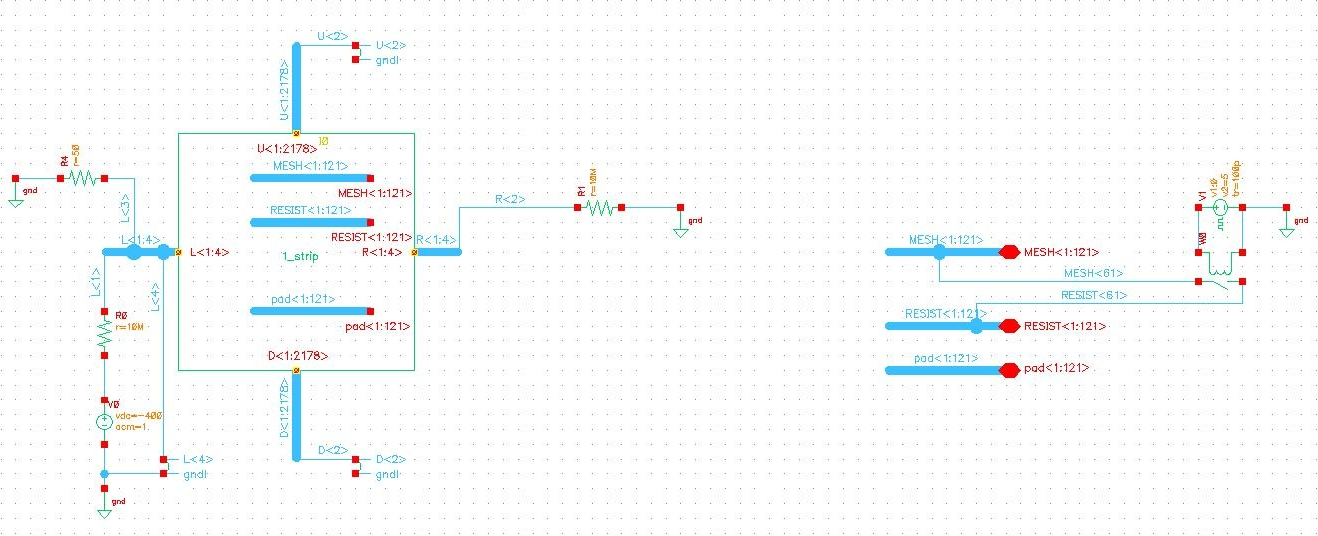}\caption{\label{fig4:Simulation_1_Strip}Single strip simulation diagram}

\par\end{centering}

\end{figure}

\par\end{center}

The discharge is produced by a switch that connect the mesh and the
resistive layer with the following parameters (opening time = closing
time = 100 ps and the pulse duration = 1 ns). One can notice on the
diagram that the resistive strip is connected to ground with a $10M\Omega$
resistor and the read out strip is connected to a $50\Omega$ resistor
to modelise the input of the charge preamplifier. The voltage applied
on the mesh is -400 V to simulate the difference of potential applied
between the mesh and the resistive strip in the real case.

\section{Results}

As it has been previously mentioned, we have simulated 3 different
sizes of strip widths. We will present the results obtained for each
configurations using the model presented before.

\subsection{Strip width of $100\mu m$}

We present the results obtained with a width of strip of $100\mu m$.
We have 9 elementary cells per element (to get a longitudinal size
of the element close to 1 mm) and 121 elements to modelise the complete
length of the strip. The elements are numbered from 1 to 121 and the
discharge is located on the middle of the strip (element \#60). 

We have performed the simulations with 4 different values of resistivity
($10k\Omega/\square$,$100k\Omega/\square$, $1M\Omega/\square$and
$10M\Omega/\square$) to observe the effect of the resistivity.

\subsubsection{Signal on the readout strip}

The figure \ref{fig5:Discharge_RO_Strip_100um} shows the maximum
voltage drop reached for the 121 elements of the readout strip.

\begin{center}
\begin{figure}[h]
\begin{centering}
\includegraphics[scale=0.2]{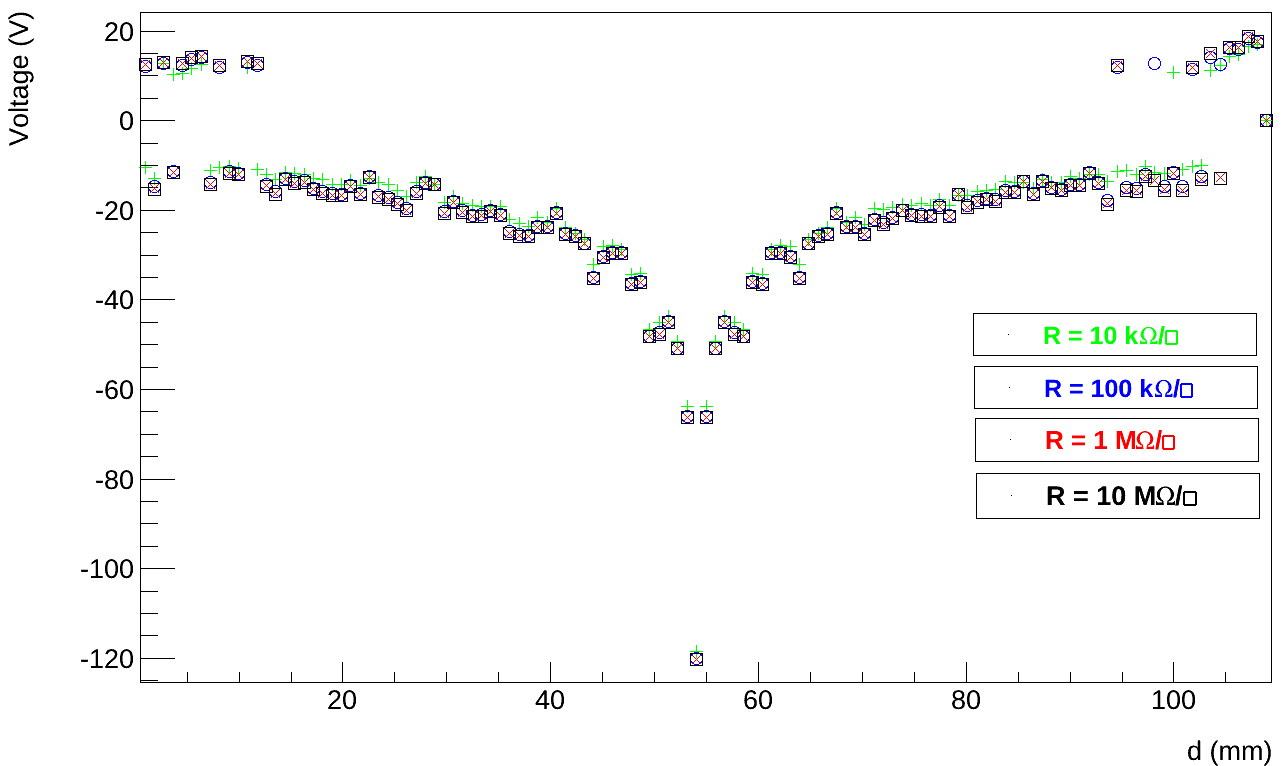}\caption{\label{fig5:Discharge_RO_Strip_100um}Maximum voltage drop reached
for each of the 121 elements of the readout strip for the four values
of resistivity}

\par\end{centering}

\end{figure}

\par\end{center}

We can notice that the element containing the discharge has the higher
voltage drop (-120 V) for the 4 different values of resistivity. We
observed a decrease of this drop for the elements placed away of the
central one. At the end of the strip, the curve shows alternatively
positive and negative values. One possible explanation of those oscillations
at the beginning and the end of the strip could be due to the inductances
introduced in our model. This plot shows that the value of the resistivity
does not seems to have an important effect on the value of the maximum
voltage drop reached by the elements of the strips, we observe a similar
shape and a similar order of magnitude of the values of the drop for
the different values of resistivity.

\subsubsection{Signal on the resistive strip}

The figure \ref{fig7:Signal_Discharge_Element_and_10_neighbors} shows
the typical signal observed on different elements of the resistive
strips. We see the voltage drops for the discharge element (\#60)
and its ten following neighbors (\#62 to \#72) versus time as provided
by the simulator at its output.

\begin{center}
\begin{figure}[!h]
\begin{centering}
\includegraphics[scale=0.15]{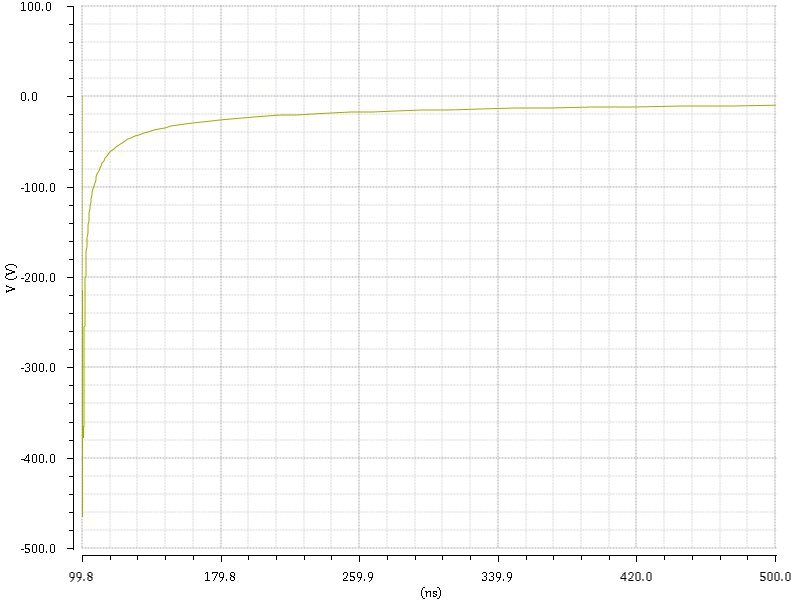}\includegraphics[scale=0.15]{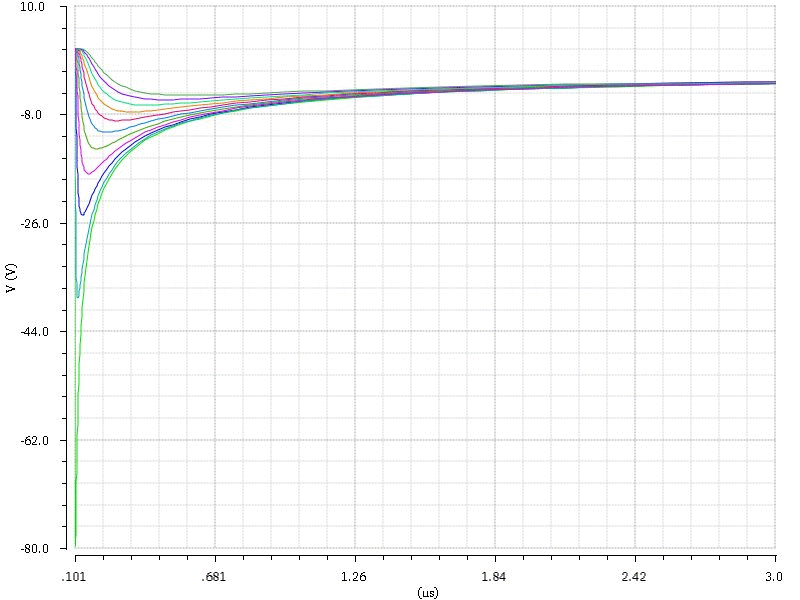}
\par\end{centering}

\begin{centering}
\includegraphics[scale=0.15]{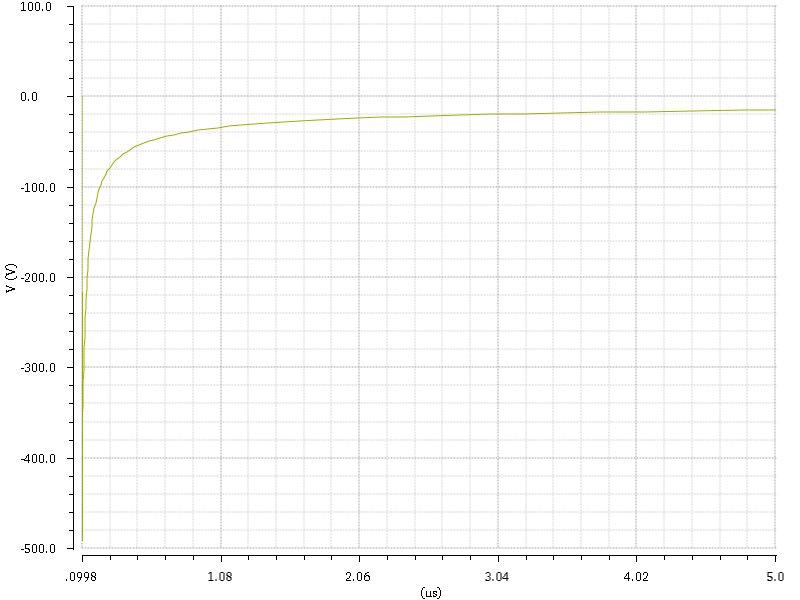}\includegraphics[scale=0.15]{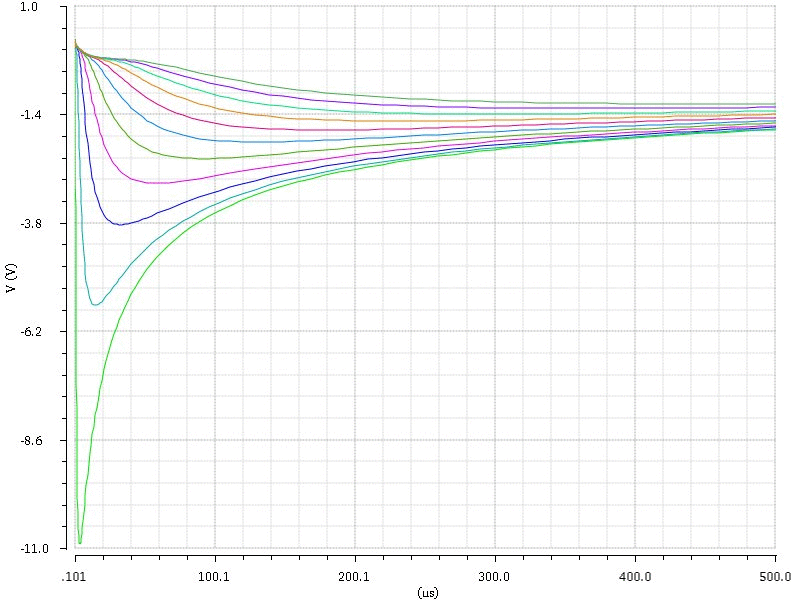}\caption{\label{fig7:Signal_Discharge_Element_and_10_neighbors}The top plots
represent the discharge element (left) and its 10 neighbors (right)
for a $10k\Omega/\square$ resistivity and the bottom plots represent
the discharge element (left) and its 10 neighbors (right) for a $10M\Omega/\square$
resistivity}

\par\end{centering}

\end{figure}

\par\end{center}

The plot with the lowest value of resistivity ($10k\Omega/\square$)
shows a higher spread of the discharge signal on the neighbors elements
(\#62 to \#72) with a higher voltage drop of these elements (few tens
of volts) but with a time propagation of $\sim 1\mu$s. This is not the case with the higher value of resistivity
($10M\Omega/\square$) where we observe a spread of the discharge
signal of only few volts on the neighbouring elements but with a longer time propagation (few hundreds of $\mu$s).

The figure \ref{fig6:Discharge_Resistive_Layer_100um} presents the
maximum voltage drop observed on the resistive layer.

\begin{center}
\begin{figure}[!h]
\begin{centering}
\includegraphics[scale=0.12]{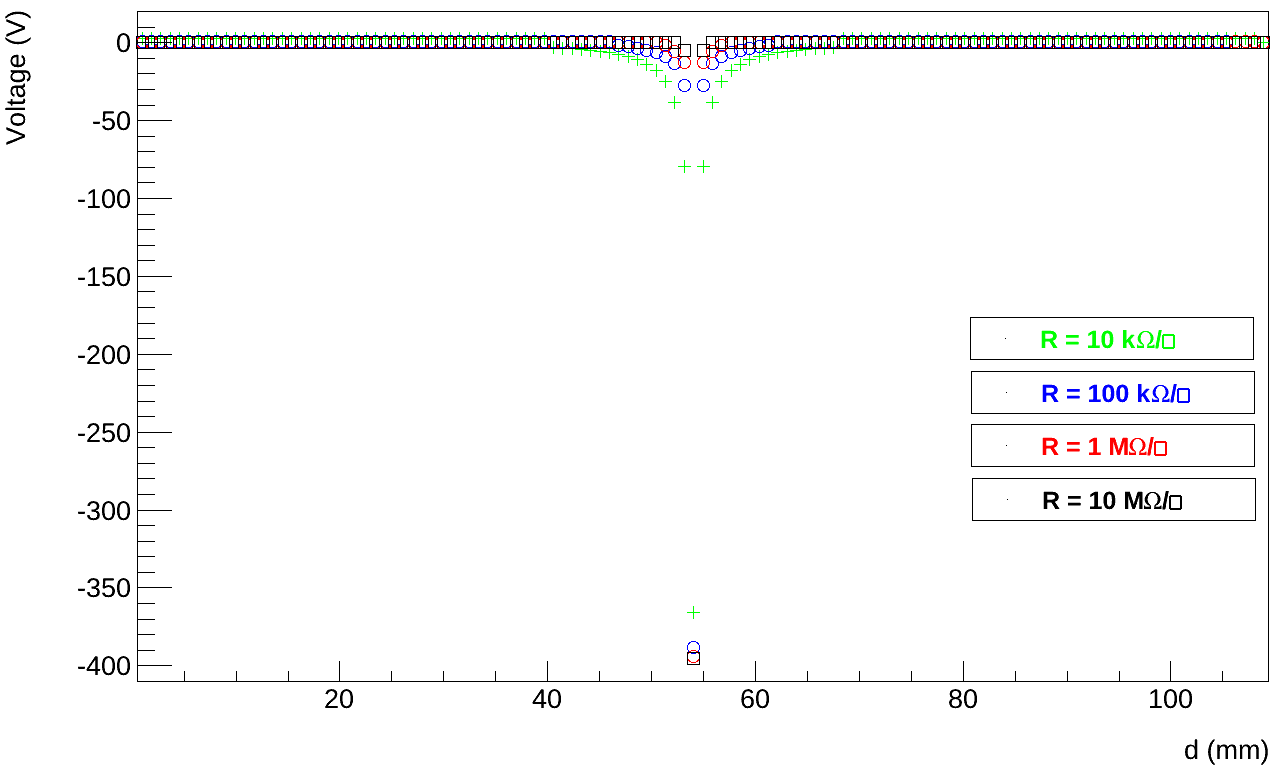}\includegraphics[scale=0.12]{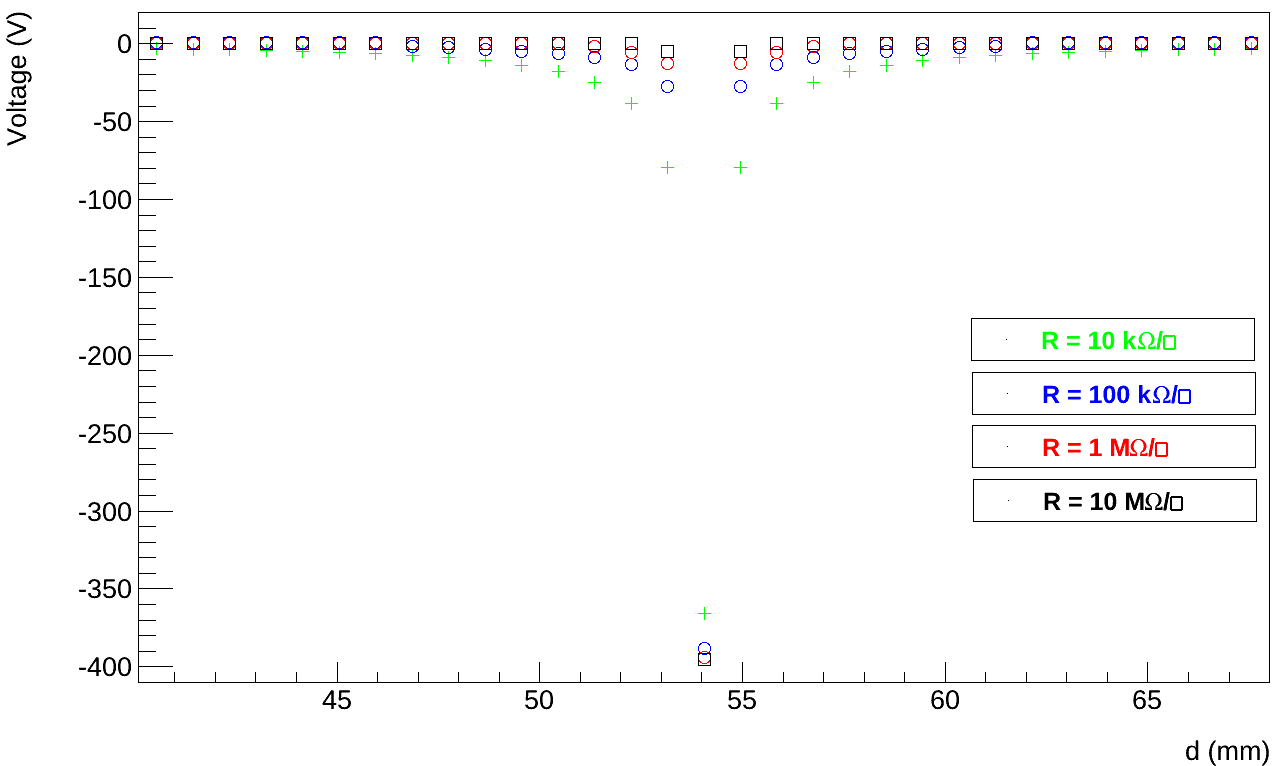}\caption{\label{fig6:Discharge_Resistive_Layer_100um}Maximum voltage drop
reached for each of the 121 elements of the resistive layer (on the
left, the full resistive strip and on the right a zoom on the central
area around the discharge)}

\par\end{centering}

\end{figure}

\par\end{center}

On this figure, we can observe the effect of the resistivity on the
signal spread on the resistive layer, the higher the resistivity,
the smaller is the spread. For the higher value of the resisitivity
($10M\Omega/\square$), we see that 3 elements have a voltage drop,
the element that contains the discharge (voltage drop = -400 V) and
its 2 closest neighbors (voltage drop $\sim$ 10 V). For the lowest
value of resistivity ($10k\Omega/\square$), we observe an area of
23 elements that have a non negligible voltage drop, the central element
that contains the discharge (voltage drop = -370 V) and the 11 closest
neighbors (voltage drop from -80 V for the 2 closest ones and -5 V
for the 2 farest ones).

\subsection{Strip width of $200\mu m$}

In this section, we present the results obtained with a width of strip
of $200\mu m$. As previously mentioned before, an element is composed
by 5 elementary cells to get a 1mm element length for a total length
of 121 mm. The location of the discharging element is the same as
before (element \#60).

The simulations were done for 4 different values of resistivity ($10k\Omega/\square$,$100k\Omega/\square$,
$1M\Omega/\square$and $10M\Omega/\square$) to observe the effect
of the resistivity.

\subsubsection{Signal on the readout strip}

The figure \ref{fig8:Discharge_RO_Strip_200um} shows the maximum
voltage drop of the 121 elements composing the readout strip.

\begin{center}
\begin{figure}[h]
\begin{centering}
\includegraphics[scale=0.2]{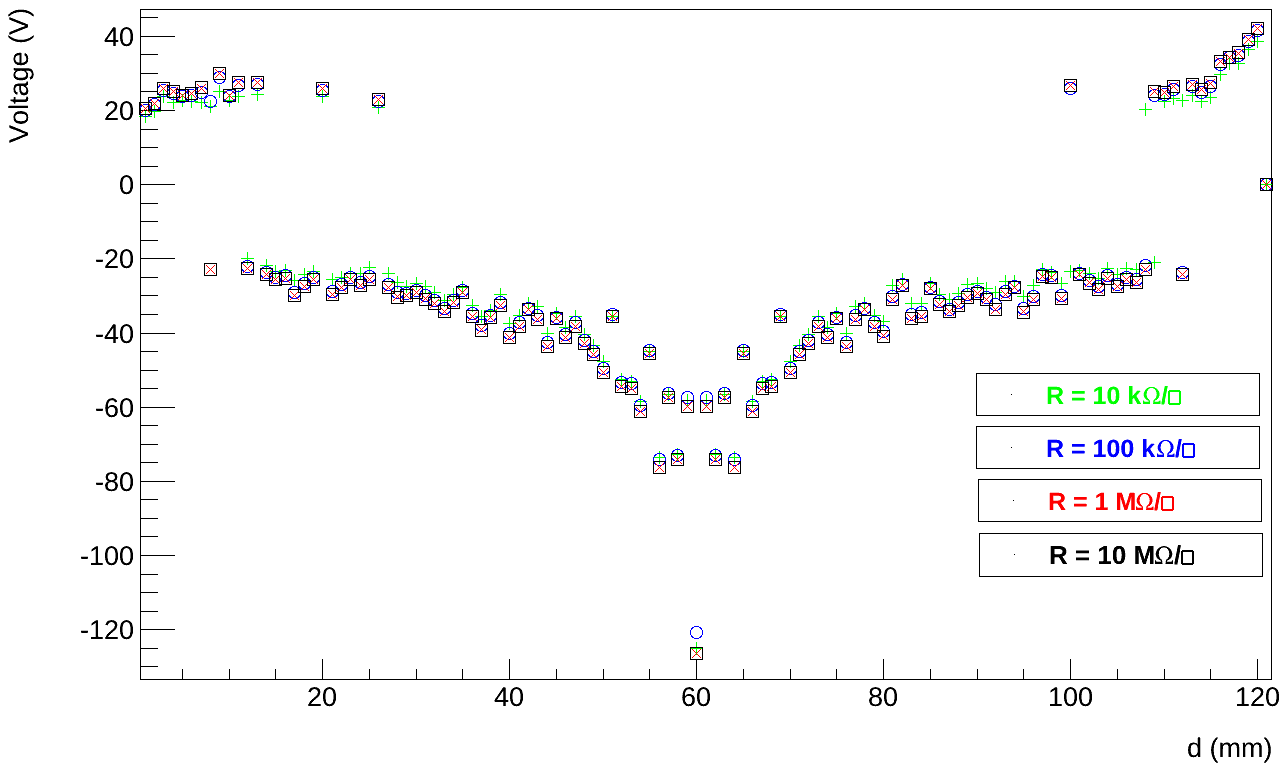}\caption{\label{fig8:Discharge_RO_Strip_200um}Maximum voltage drop reached
for each of the 121 elements of the readout strip for the four values
of resistivity}

\par\end{centering}

\end{figure}

\par\end{center}

The element containing the discharge has the higher voltage drop (-130
V) for the 4 different values of resistivity. We observed a decrease
of this drop for the elements placed away of the central one. We also
observe the same effect of oscillations at the ends of the strips.
One can notice that the values of the voltage drop reached by the
elements of the $200\mu m$ are higher than those reached with the
$100\mu m$ one. The value of the resistivity does not seem to have
an important effect on the value of the maximum voltage drop reached
by the elements of the strips, we observe a similar shape and a similar
order of magnitude of the values of the drop for the different values
of resistivity.

\subsubsection{Signal on the resistive strip}

We observe the results of the voltage drop reached by the 121 elements
of the resistive strip on figure \ref{fig9:Discharge_Resistive_Layer_200um}.

\begin{center}
\begin{figure}[h]
\begin{centering}
\includegraphics[scale=0.12]{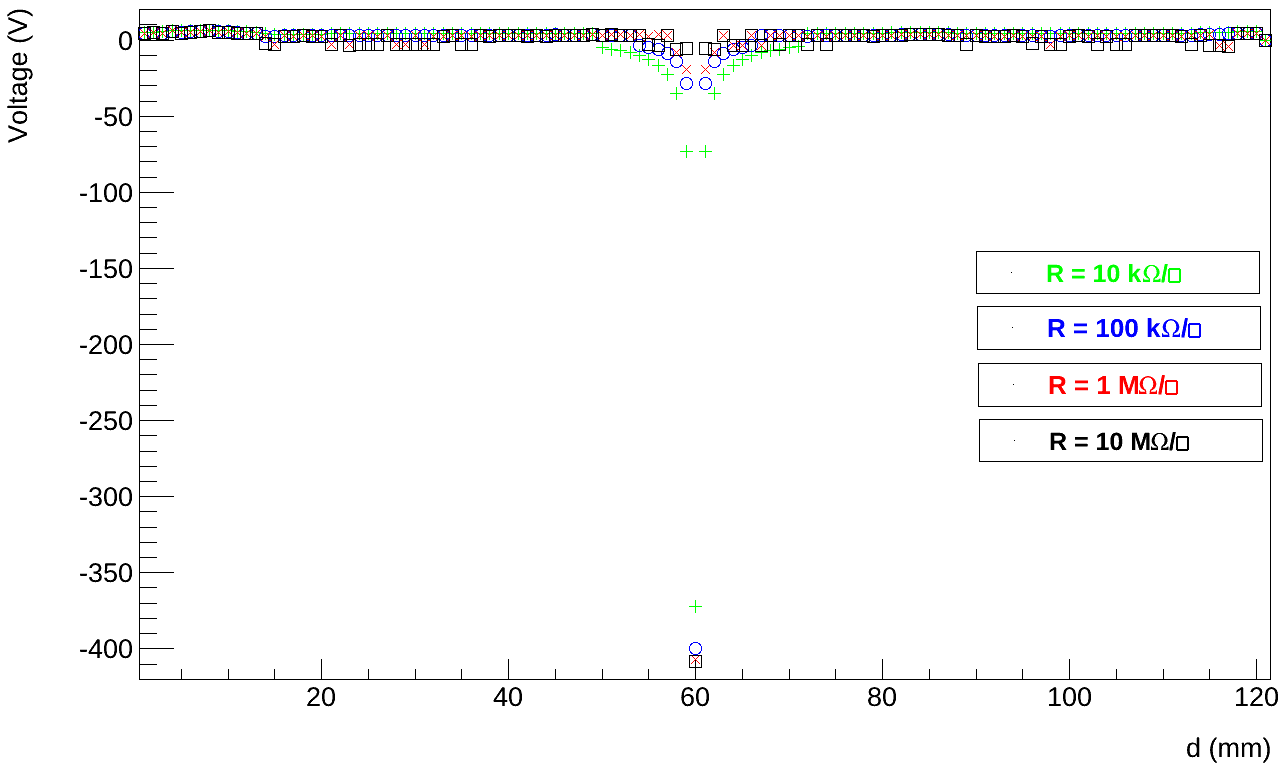}\includegraphics[scale=0.12]{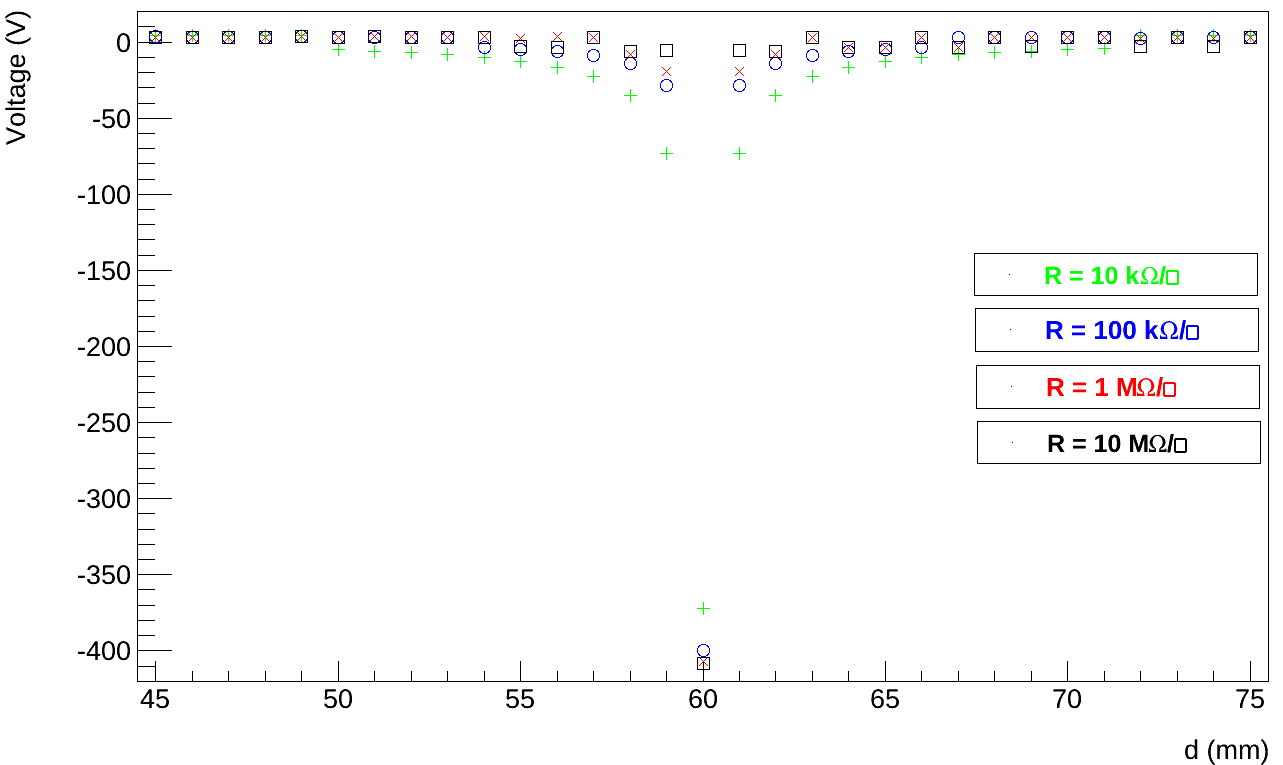}\caption{\label{fig9:Discharge_Resistive_Layer_200um}Maximum voltage drop
reached for each of the 121 elements of the resistive layer (on the
left, the full resistive strip and on the right a zoom on the central
area around the discharge)}

\par\end{centering}

\end{figure}

\par\end{center}

This figure shows the effect of the value resistivity on the signal
spread on the resistive layer. For the higher value ($10M\Omega/\square$)
of the resistivity, the spread is smaller. We observe 10 elements
(except the one containing the discharge) that have a noticeable voltage
drop (5V), the element that contains the discharge (voltage drop =
-410 V). For the lowest value of resistivity ($10k\Omega/\square$),
we observe an area of 22 elements (except the discharging one) that
have a non negligible voltage drop (10 V or more), the central element
that contains the discharge (voltage drop = -370 V).

\subsection{Strip width of $300\mu m$}

In this section, we present the results obtained with a width of strip
of $300\mu m$. We have 3 elementary cells per element (to get a longitudinal
size of the element close to 1 mm) and the others parameters are the
same as the ones used for the two widths of strips simulated before
(number of elements for one strip, element that contains the discharge
and number and value of resistivity simulated).

\subsubsection{Signal on the readout strip}

The maximum voltage drop of elements of the strip is showed on figure
\ref{fig10:Discharge_RO_Strip_300um}.

\begin{center}
\begin{figure}[h]
\begin{centering}
\includegraphics[scale=0.2]{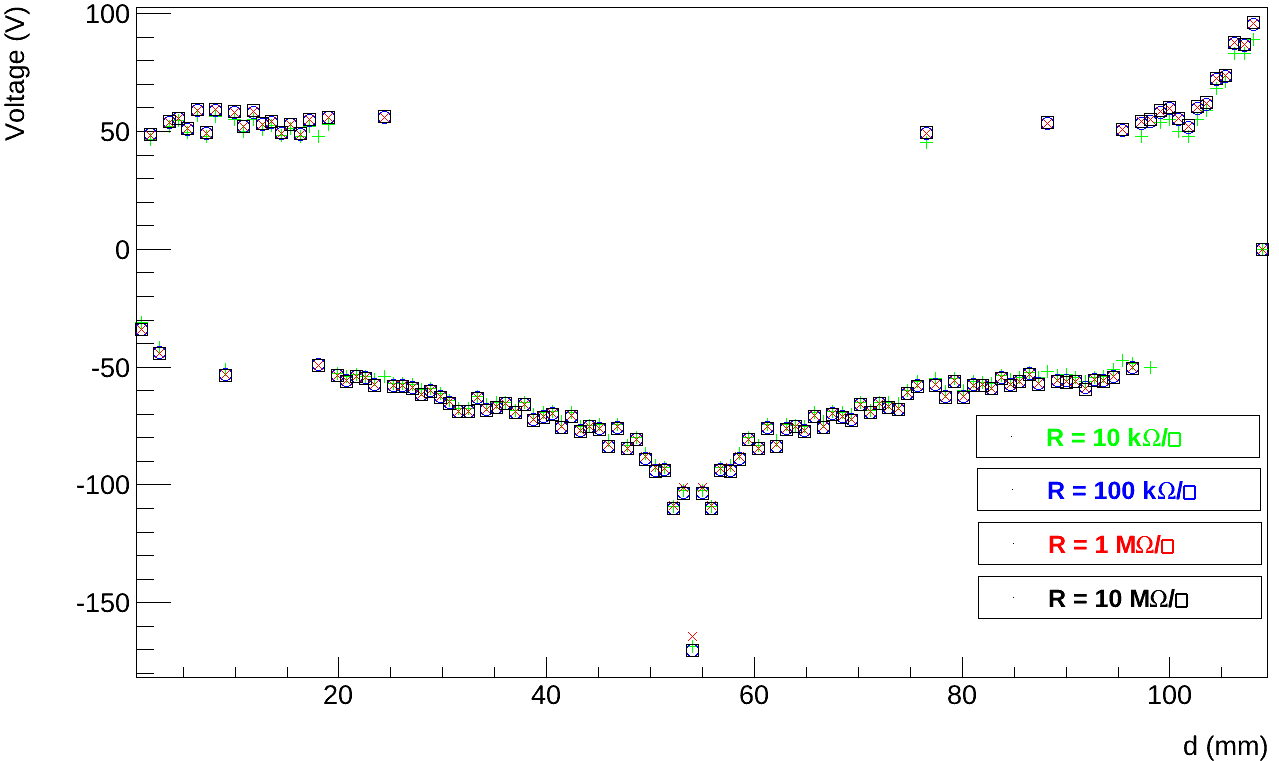}\caption{\label{fig10:Discharge_RO_Strip_300um}Maximum voltage drop reached
for each of the 121 elements of the readout strip for the four values
of resistivity}

\par\end{centering}

\end{figure}

\par\end{center}

The element that contains the discharge (\#60) shows the higher voltage
drop (-170V) for all the values of resistivity. The voltage drop decreases
for the elements along the strips ($-120V$ for the close neighbors
to $\pm50V$ for the elements at the ends of the strip). We still
observe oscillations at the ends of the strip. One can also notice
that the value of the resisitivity does not play a role on the voltage
drop of the elements, the shape and the magnitude remaining similar
for the 4 values of resistivity.

\subsubsection{Signal on the resistive strip}

The results concerning the resistive elements are presented on figure
\ref{fig11:Discharge_Resistive_Strip_300um}.

\begin{center}
\begin{figure}[h]
\begin{centering}
\includegraphics[scale=0.12]{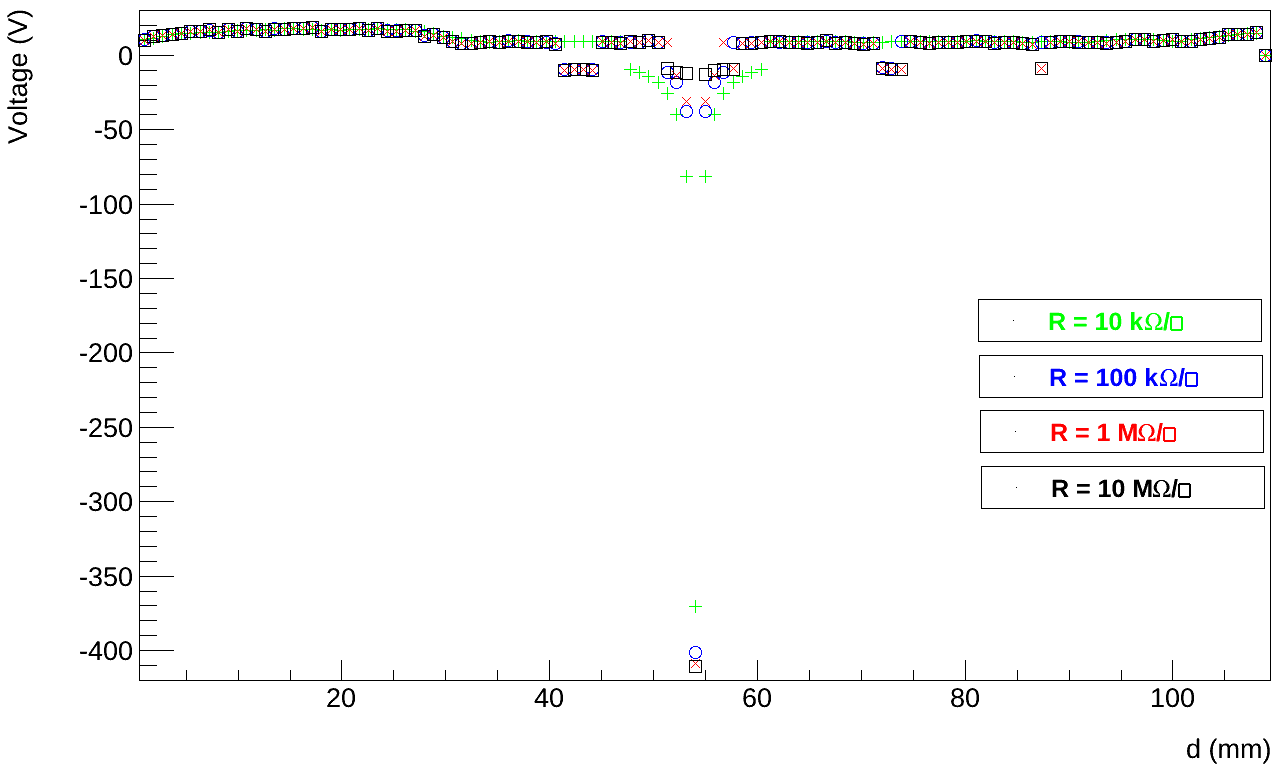}\includegraphics[scale=0.12]{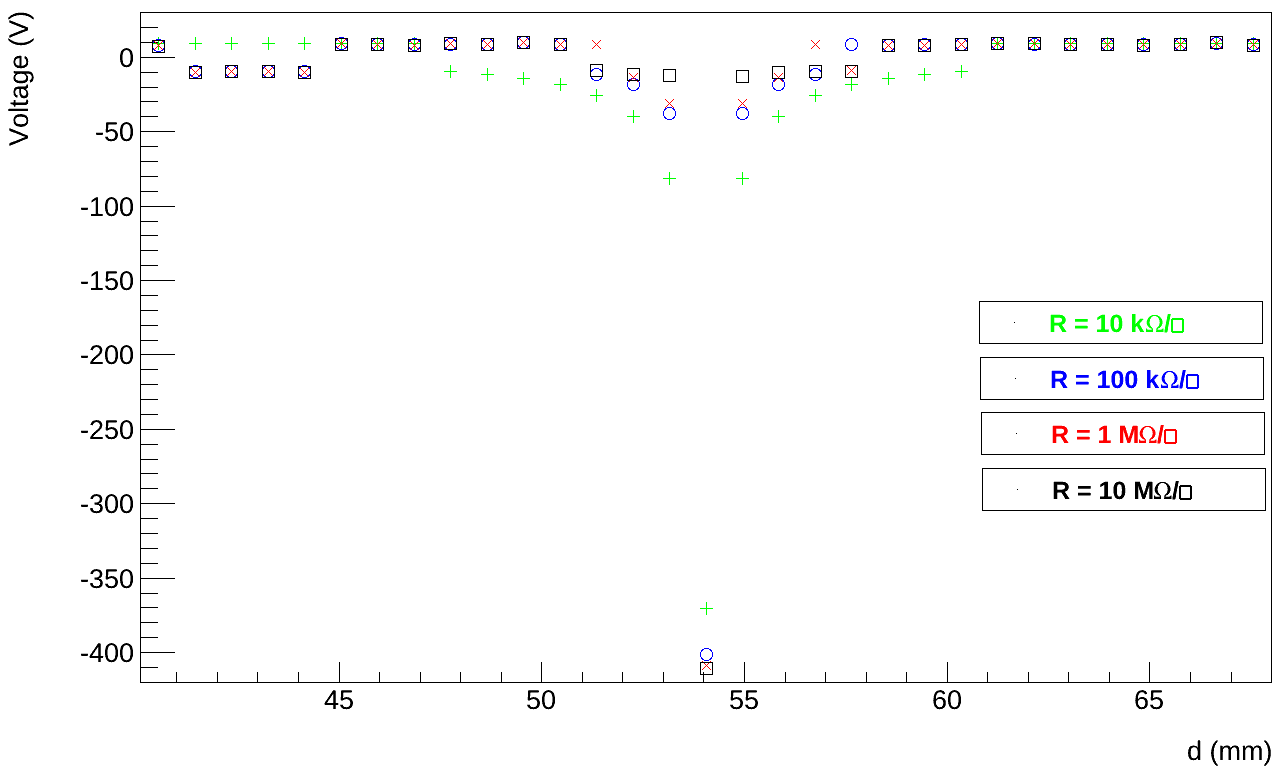}\caption{\label{fig11:Discharge_Resistive_Strip_300um}Maximum voltage drop
reached for each of the 121 elements of the resistive layer (on the
left, the full resistive strip and on the right a zoom on the central
area around the discharge)}

\par\end{centering}

\end{figure}

\par\end{center}

This figure shows the effect of the value of the resistivity on the
signal spread on the resistive layer. For the higher value ($10M\Omega/\square$)
of the resistivity, the voltage drop of the discharging element is
higher (-410V) where with the lowest value of resisitivity ($10k\Omega/\square$),
we observe a smaller voltage drop of the same element (-370V). Concerning
the spread of the discharge, in case of the lowest value of resistivity,
the area involved in the discharge is 7 elements around the discharging
one with a maximum voltage drop value from -100V to -20V. For the
highest value of resistivity, the area involved in the discharge is
smaller (4 elements around the discharging one) with value of maximum
voltage drop of -10V. the other elements have a maximum voltage drop
of +10V.

\subsection{Effect of the inductance values for the mesh electrode and the readout
strip}

In this section, we have tested the effect of the inductance value
for the mesh and the read strip for a $200\mu m$ strip width keeping
constant all the other parameters of the configuration.

\subsubsection{Effect of the pad inductance value}

We present in this part, the effect of the pad inductance value used
in our model of simulation. We have simulated the single strip geometry
presented in fig.\ref{fig4:Simulation_1_Strip} with 4 different values
of pad inductance (0 pH, 10 pH, 41.5 pH and 83 pH) and a constant
value of the mesh inductance (120 pH). The result is presented in
figure\ref{fig12:Effect of the pad inductance value}.

\begin{center}
\begin{figure}[h]
\begin{centering}
\includegraphics[scale=0.2]{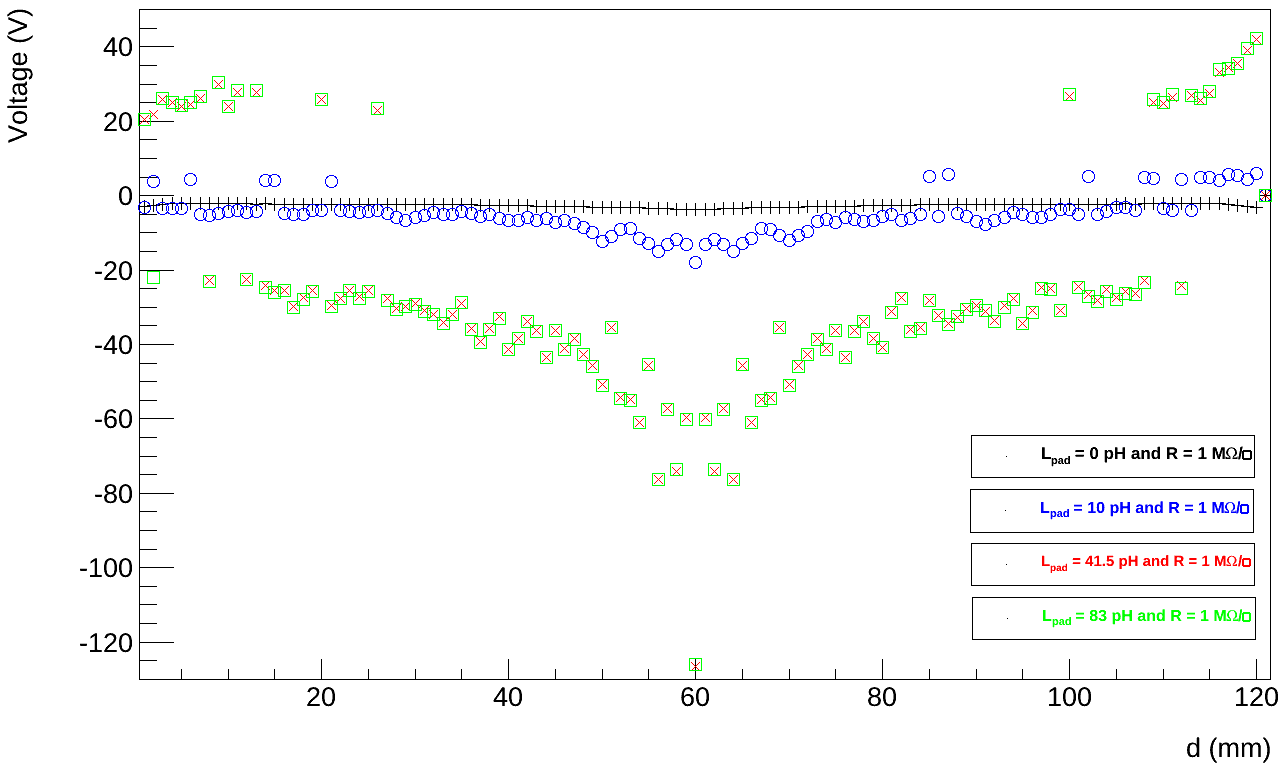}\caption{\label{fig12:Effect of the pad inductance value}Maximum voltage drop
reached by the 121 elements of the readout strip for four values of
pad inductance}

\par\end{centering}

\end{figure}

\par\end{center}

The plot shows the effect of the inductance value. The value previously
used in our model for a $200\mu m$ strip width is 83 pH. Table \ref{tab3:Voltage drop vs pads inductance value}
presents the value of the maximum voltage drop reach by the central
element (\#60) and lowest voltage drop reach by elements located at
both ends of the strip.

\begin{center}
\begin{table}[h]
\begin{centering}
\begin{tabular}{|c|c|c|}
\hline 
$L_{pad}$ (pH) & Voltage drop 1 (V) & Voltage drop 2 (V)\tabularnewline
\hline 
\hline 
0 & $-3.8\times10^{-3}$ & $-2\times10^{-3}$\tabularnewline
\hline 
10 & -18 & $\pm$5\tabularnewline
\hline 
41.5 & -130 & $\pm$25\tabularnewline
\hline 
83 & -130 & $\pm$25\tabularnewline
\hline 
\end{tabular}\caption{\label{tab3:Voltage drop vs pads inductance value}Maximum value of
voltage drop for the central element (\#60) (voltage drop 1) and at
the ends of strip (voltage drop 2) as a function of the pad inductance
value}

\par\end{centering}

\end{table}

\par\end{center}

One can notice that the pad inductance plays an important role in
the voltage drop reach by each of the 121 elements composing the strip.
Without any inductance of the strip (0 pF), the voltage drop reaches
a value of few mV. As soon as this value increases to 10 pF, the voltage
drop increases to -18 V for the element \#60 and $\pm$ 5V for the
elements located at the ends of the strip. If this value still increase
to 41.5 pF, we observe high value of voltage drop for the central
element (-130 V) and $\pm25V$ for elements located at the end of
the strip. We can also notice that we do not observe any difference
in the value of the voltage drop obtained with 41.5 pH and 83 pH and
also in the shape of the plot.

\subsubsection{Effect of the mesh inductance value}

In this part, we present the value of the maximum voltage drop reached
by the 121 elements of the readout strip for different values of the
mesh inductance (0, 60, 120 and 240 pH). Those simulations have been
realised for a width of strips of $200\mu m$ and the results are
presented in figure \ref{fig13:Mesh_Inductance_Value_Effect}.

\begin{center}
\begin{figure}[h]
\begin{centering}
\includegraphics[scale=0.2]{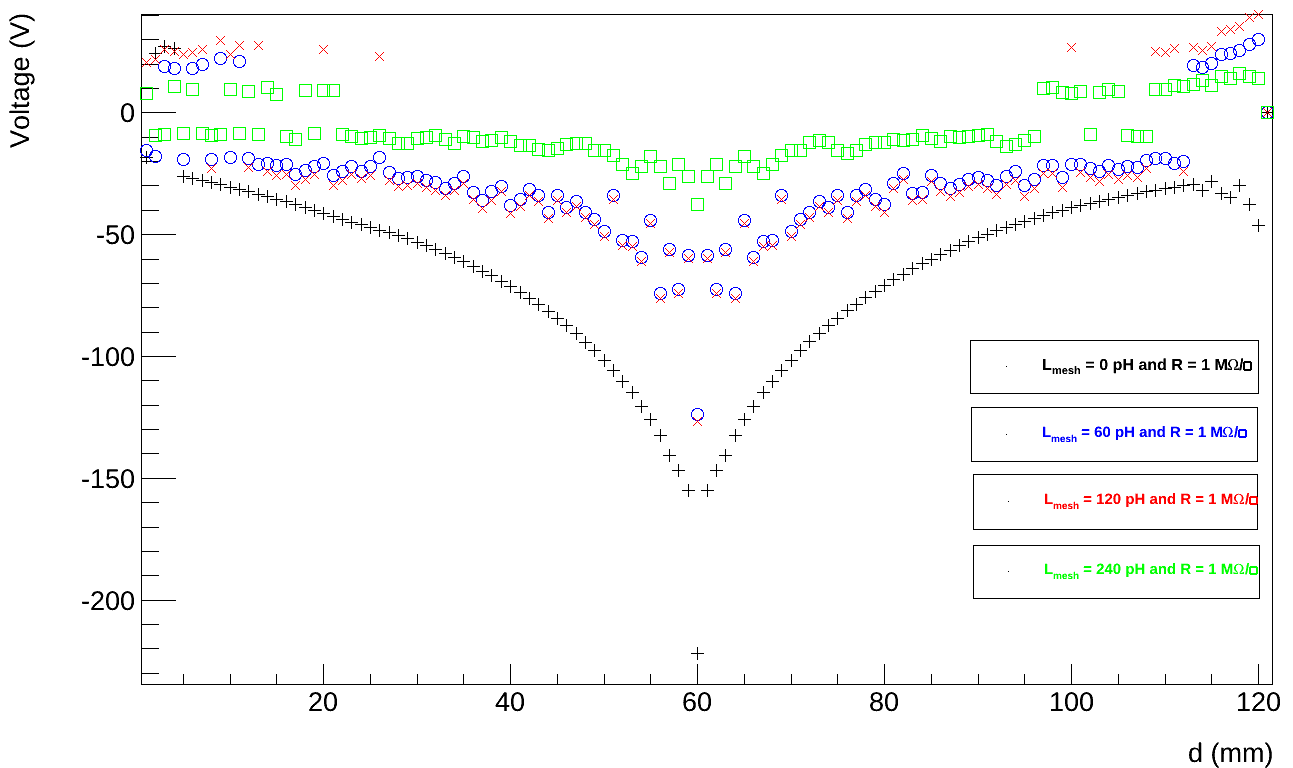}
\par\end{centering}

\caption{\label{fig13:Mesh_Inductance_Value_Effect}Maximum voltage drop reached
by the 121 elements of the readout strip for four values of mesh inductance}

\end{figure}

\par\end{center}

We can observe that the higher is the value of the value of the mesh
inductance, the lower is the voltage drop of the elements composing
the readout strip. The value of the voltage drop are summarized in
table \ref{tab4: voltage drop vs mesh inductance}.

\begin{center}
\begin{table}[h]
\begin{centering}
\begin{tabular}{|c|c|c|}
\hline 
$L_{mesh}$ (pH) & Voltage drop 1 (V) & Voltage drop 2 (V)\tabularnewline
\hline 
\hline 
0 & -221 & $\pm$27\tabularnewline
\hline 
60 & -124 & $\pm$20\tabularnewline
\hline 
120 & -127 & $\pm$20\tabularnewline
\hline 
240 & -38 & $\pm$10\tabularnewline
\hline 
\end{tabular}
\par\end{centering}

\caption{\label{tab4: voltage drop vs mesh inductance}Maximum value of voltage
drop for the central element (\#60) (voltage drop 1) and at the ends
of strip (voltage drop 2) as a function of the mesh inductance value}
\end{table}

\par\end{center}

The results presented in table \ref{tab4: voltage drop vs mesh inductance}
show clearly see the importance of the value of the mesh inductance.
There is more than a factor five between the voltage drop reaches
without any inductance for the central element and with the highest
value of mesh inductance for the same element (-221 V for 0 pH and
-38 V for 240 pH). We observe the same trend for the ends of the strip
with a lower magnitude ($\pm$27 for 0 pH and $\pm$10 V for 240 pH).
The last point that one can notice is that with a value of $L_{mesh}$
of 0 pH, the rebounds, present with all the others values of inductance,
are suppressed.

\subsubsection{Results without any inductance in the model}

In this section, we present the results of the simulation obtained
with a model without any inductance ($L_{mesh}=L_{pad}=0$ pH). The
results of the maximum voltage drop of the elements composing the
readout strip are presented in figure \ref{fig14: Pads voltage drop with Lmesh =00003D Lpads  =00003D 0 pH}.

\begin{center}
\begin{figure}[h]
\begin{centering}
\includegraphics[scale=0.2]{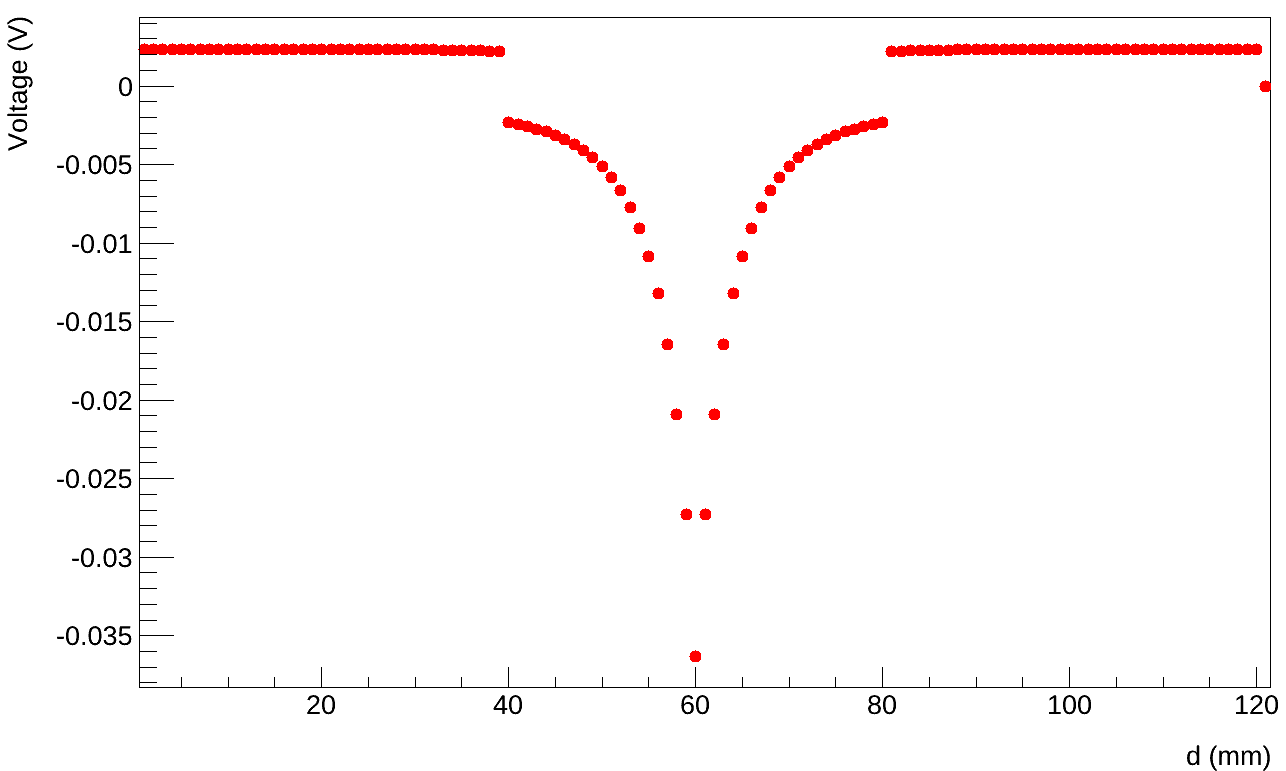}\caption{\label{fig14: Pads voltage drop with Lmesh =00003D Lpads  =00003D 0 pH}Maximum
voltage drop reached by the 121 elements of the readout strip without
any inductance}

\par\end{centering}

\end{figure}

\par\end{center}

Without any inductances, the maximum voltage drop reached by the pads
is very low (-36 mV for the central element and +2 mV for the elements
located at both ends of the strip). One can notice that the results
obtained without any inductance is 3 orders of magnitude lower than
those obtained with the computed inductance values ($L_{mesh}=120$
pH and $L_{pads}=83$ pH). We also observed a smooth curve without
rebounds but with two discontinuities for pad \#40 and \#80.

\section{Conclusion}

\subsection{Readout strip}

Comparing the 3 strip widths, we observe a similar shape of curve
for the 4 different values of resisitivity, nevertheless, the value
of the maximum voltage of the element of the strip increases with
the width of the strips as it is shown in table \ref{tab1:Valeur_Max_Voltage_Drop_RO_Strip}.

\begin{center}
\begin{table}[h]
\begin{centering}
\begin{tabular}{|c|c|c|}
\hline 
Width ($\mu m$) & Voltage drop 1 (V) & Voltage drop 2 (V)\tabularnewline
\hline 
\hline 
100 & -120 & $\sim\pm15$\tabularnewline
\hline 
200 & -130 & $\sim\pm25$\tabularnewline
\hline 
300 & -170 & $\sim\pm50$\tabularnewline
\hline 
\end{tabular}\caption{\label{tab1:Valeur_Max_Voltage_Drop_RO_Strip}Maximum value of voltage
drop for the central element (\#60) (voltage drop 1) and at the end
of strip (voltage drop 2) function the width of the strip }

\par\end{centering}

\end{table}

\par\end{center}

One can also notice that the larger is the strip, the larger is the
number of elements involved in the discharge (spread of the discharge
signal). This can be explained by the fact that for the $100\mu m$
width strip, an element is composed by 9 elementary cell of size $100\times100\mu m^{2}$
of a given resistivity. For the same resistivity, a $300\mu m$ width
strip will be composed by only 3 elementary cells of size $300\times300\mu m^{2}$.
The resistivity will be 9 times lower in the latter case and this
will lead to a higher number of elements involved in the discharge
area.

Another important point is the value of the inductance of the pad
used in our model. Table \ref{tab3:Voltage drop vs pads inductance value}
shows that there is a clear trade off between the case with no inductance
and the presence of, even, a small inductance (few pH), the value
of the maximum voltage drop varying by 3 orders of magnitude from
few mV to -20 V for the central element. If the value of the inductance
increases to 41.5 pH, we observe an increase of the maximum voltage
drop from -20 V to -130 V for the central element. After that, it
seems that the curve reaches a plateau, we do not observe any increase
between 41.5 pH and 83 pH.

\subsection{Resistive strip}

For each width of the strip, we have simulated 4 different values
of resistivity. We observe that the higher is the value of the resistivity,
the higher is the value of the maximum voltage drop of the central
element (\#60) and this observation is the same for the other 120
elements of the strip. The results of the maximum voltage drop for
the element containing the discharge and the spread of the discharge
signal are presented in table \ref{tab2:Value_Max_Voltage_Drop_Resistive_Strip}.

\begin{center}
\begin{table}[h]
\begin{centering}
\begin{tabular}{|c|c|c|c|c|}
\hline 
Width ($\mu m$) & $\rho$ ($\Omega/\square$) & Voltage drop 1 (V) & Spread & Voltage drop 2 (V)\tabularnewline
\hline 
\hline 
100 & 10k & -370 & $\pm$ 10 & 0\tabularnewline
\hline 
 & 100k & -390 & $\pm$ 7 & 0\tabularnewline
\hline 
 & 1M & -400 & $\pm$ 3 & 0\tabularnewline
\hline 
 & 10M & -400 & $\pm$ 1 & 0\tabularnewline
\hline 
200 & 10k & -370 & $\pm$ 11 & 5\tabularnewline
\hline 
 & 100k & -400 & $\pm$ 7 & 5\tabularnewline
\hline 
 & 1M & -410 & $\pm$ 3 & 5\tabularnewline
\hline 
 & 10M & -410 & $\pm$ 3 & 5\tabularnewline
\hline 
300 & 10k & -370 & $\pm$ 7 & 10-15\tabularnewline
\hline 
 & 100k & -400 & $\pm$ 3 & 10-15\tabularnewline
\hline 
 & 1M & -410 & $\pm$ 3 & 10-15\tabularnewline
\hline 
 & 10M & -410 & $\pm$ 3 & 10-15\tabularnewline
\hline 
\end{tabular}\caption{\label{tab2:Value_Max_Voltage_Drop_Resistive_Strip}Maximum value
of voltage drop for the central element (\#60) (voltage drop 1), the
spread of the discharge around the central element (in number of elements)
and the maximum voltage drop at the end of the strip (voltage drop
2) for the different strip widths and resistivity ($\rho$)}

\par\end{centering}

\end{table}

\par\end{center}

One can notice on figure \ref{fig6:Discharge_Resistive_Layer_100um}
that for a $100\mu m$ width strip there is a signal peak centered
on the element \#60 and the elements not involved in the discharge
have a voltage drop of 0V. This pattern is very well defined. In the
case of larger strip ($200\mu m$ and $300\mu m$), we still have
a peak due to the discharge but we also have a tail of this peak that
includes elements far away from the discharge (figures \ref{fig9:Discharge_Resistive_Layer_200um}
and \ref{fig11:Discharge_Resistive_Strip_300um}). The amplitude of
the voltage drop of the elements included in this tail increases with
the width of the strip. We observe for a $200\mu m$ width, a maximum
voltage drop at the end of the strip of $\sim$5V.

For a $300\mu m$ width, the amplitude of the voltage drop at the
end of the strip is $\sim$ 10 - 15V. We also can see on the figures
\ref{fig9:Discharge_Resistive_Layer_200um} and \ref{fig11:Discharge_Resistive_Strip_300um}
the apparition of oscillations between close elements that do not
have similar voltage drop as we could expect.

\end{document}